\documentclass[a4paper,twoside,11pt]{article}
\usepackage[utf8]{inputenc}

\usepackage[english]{babel}
\usepackage{subeqnarray}
\usepackage{url}

\usepackage{graphicx,psfrag}
\usepackage{fixltx2e,amsmath}
\usepackage{xcolor}
\usepackage{amssymb}
\usepackage[final]{pdfpages}
\usepackage{bbold}
\usepackage{dsfont}
\usepackage{pgfplots}
\usepackage{subfig}
\usepackage{algorithm}
\usepackage{algorithmic}
\usepackage[font=small,labelfont=bf]{caption}

\usepackage[font={footnotesize}]{caption}

\usepackage[utf8]{inputenc}
\usepackage[T1]{fontenc}
\usepackage{xcolor,amsmath,amssymb}
\usepackage{geometry}
\usepackage{graphicx}
\usepackage[sort&compress,numbers]{natbib}
\usepackage{enumerate}
\usepackage{wasysym}
\usepackage[textsize=footnotesize]{todonotes}
\usepackage{url}

\usepackage{pgfplots}

\definecolor{darkblue}{rgb}{0,0,0.6}
\definecolor{darkred}{rgb}{0.6,0,0}

\usepackage{hyperref}

\usepackage{authblk}

\usepackage[bitstream-charter]{mathdesign}

\usetikzlibrary{pgfplots.groupplots}

\graphicspath{{figures/}} 

\title{How Much Income Inequality Is Too Much?}

\author{Jean-Philippe Bouchaud}
\affil{Capital Fund Management \& Acad\'emie des Sciences, Paris} 
\begin{document}

\date{\today}

\maketitle

\begin{abstract}
We propose a highly schematic economic model in which, in some cases, wage inequalities lead to higher overall social welfare. This is due to the fact that high earners can consume low productivity, non essential products, which allows everybody to remain employed even when the productivity of essential goods is high and producing them does not require everybody to work. We derive a relation between heterogeneities in technologies and the minimum Gini coefficient required to maximize global welfare. Stronger inequalities appear to be economically unjustified. Our model may shed light on the role of non-essential goods in the economy, a topical issue when thinking about the post-Covid-19 world.
\end{abstract}

\section{Introduction}

The issue of income and wealth inequalities is now at the top of the political agenda in many countries. It is also a fascinating scientific problem, with a flurry of academic studies, ranging from detailed data analysis to economic models, historical accounts and philosophical arguments. Piketty's ``Capital in the XXIst Century'' is a worldwide bestseller \cite{Piketty1}, maybe soon joined by his second book, ``Capital \& Ideology'' \cite{Piketty2}. Statistical models of income and wealth dynamics have a long history, starting with Champernowne \cite{Champ} and Angle \cite{Angle}, with a particular upsurge in the ``Econophysics'' literature since 2000 -- for recent reviews see e.g. \cite{chakraborti,boghossian,brazil}, and, for economics papers \cite{Benhabib,Moll}.

Whereas too much inequality (as well as too much equality) seems both morally unfair and economically inefficient, there is no compelling argument allowing one to decide which level of inequality policy makers should target. If one thinks, for example, in terms of the classic Gini coefficient\footnote{The Gini coefficient $G \in [0,1]$ measures the average absolute income (or wealth) difference in a population, rescaled by (twice) the average wealth.} $G$, is there an optimal value of $G$ that we can rationally agree on? Should $G$ be less than $0.3$ like in Scandinavian countries, or is $0.35$ or even $0.4$ still acceptable? Should this target be independent of the level of development of an economy? How does the level of inequality affect economic growth and global social welfare \cite{imf}? Are ``the rich'' catalysts or parasites?

These questions are unfortunately fraught with ideological considerations. The rich are depicted as heroes or villain in a knee-jerk fashion, depending on which side of the political spectrum one speaks from. Is there a way to frame the debate in more neutral, rational terms? John Rawles, for example, famously argued that the income distribution should be such that the wage of the poorest households is maximized \cite{Rawles}. This, however, does not {\it a priori} constrain the Gini index in any way. Free market advocates believe that the whole debate is moot: whatever comes out of a free market economy is by definition optimal \cite{Mankiw}. There is a growing literature on optimal taxes, in particular progressive, redistributive taxes and/or bequeath taxes, see e.g. \cite{Piketty,bouchaud,violante}. A standard argument for limiting tax progressivity (and therefore keeping inequalities above a certain level) is a potential reduction of the incentives to work and to invest in skills. A stronger public sector, on the other hand, can be beneficial in terms of long term growth, through different channels such as health, education, public research, etc. \cite{mazzucato}. However, crazy-looking initiatives launched by single individuals, which would never be decided collectively by any sensible public agency, can lead to ground-breaking innovations that are collectively beneficial.   

Another interesting line of thought was developed by V. Venkatasubramanian in a recent book called ``How Much Inequality Is Fair?'' \cite{venka}, transposing ideas from statistical mechanics (in particular the maximum entropy criterion) to define an acceptable level of inequalities. 

The aim of this note is to explore another avenue, namely the impact of economic development and labour productivity on wage inequalities. We argue that when the utility associated to the consumption of goods saturates (i.e. when further consumption brings no extra utility at all), then increased productivity can lead to unemployment, which is an extreme form of income inequality. For full employment to be maintained, the consumption of low productivity, expensive goods must be bolstered. In some cases, which we explore below within a highly simplified model, this is only possible when some households earn more than others. We derive the minimum value of the Gini coefficient for full employment (and hence maximum total utility) to be achieved. In a nutshell, our argument is based on the idea that if the productivity of a non-essential good is too low, its price is too high for any agent to afford in a strictly egalitarian society. Hence, this good is not produced at all. But if at the same time the essential good is easy to produce while its utility saturates, only a fraction of the labour force is needed and output remain low. In this case, global welfare is higher when the Gini coefficient is non-zero, i.e. in the presence of some degree of wage inequalities. These can be however partly compensated by some redistributive policies, since tax revenues can be increased.

\section{Homogeneous Agents, Equal Wages}

We imagine an economy that can produce two goods, 1 \& 2, with linear technologies. Production of good $i$ is given by
\begin{equation}
Y_i = z_i N_i,
\end{equation}
where $z_i$ is productivity and $N_i$ labour. Assuming market clearing and, for now, identical wages $w$ for all, zero profit condition gives the prices at which goods should sell:\footnote{Of course, the average wage can always be taken as the unit in which prices are counted, but we keep $w$ explicit for clarity.}
\begin{equation}
p_i = \frac{w}{z_i}.
\end{equation}

We assume $z_1 > z_2$ and $p_1 < p_2$: Good 1 is a bare necessity good (e.g. wheat), so technology has evolved to make it easy to produce, whereas good 2 is a non-essential, ``luxury'' good (e.g. art), with low productivity. Good 1 is thus cheap, and good 2 is pricey.  

In order to keep the algebra as simple as possible, we consider households with {\it strictly saturating utility functions} for both goods:
\begin{equation} \label{eqU}
U = \sum_i \theta_i \min[c_i,\Gamma_i],
\end{equation}
where $\Gamma_i$ is the consumption level beyond which no further utility is drawn from consumption. As Marcel Dassault once quipped, one cannot eat more than one chicken a day; in some cases one can even imagine that utility starts decreasing beyond some threshold where more goods become useless and expensive to store. Since we assume good 1 is vital, we will set $\theta_1 > \theta_2$, i.e. households care much more about buying good 1 and only start buying good 2 when they have quenched their thirst for good 1.

The strict saturation of the utility function is important but not essential in what follows. For non saturating utility functions, there is no involuntary unemployment even when the productivity is large, since households are always happy to consume more goods -- see Appendix \ref{sec:appendix} for a discussion of such cases. As we will show below, this is not true anymore when $U$ saturates, as in Eq. \ref{eqU}.

Finally, we make the crucial assumption that goods are not infinitely divisible, i.e. each agent cannot consume less than a ``quantum'' of good $\gamma_i$. There is little use in buying half a painting, or half a personal computer or half a meal in a posh restaurant, among many examples of expensive, non essential and indivisible goods. Hence $c_i \geq \gamma_i$ for all agents who consume good $i$. Clearly, one must impose $\Gamma_i > \gamma_i$ for the model to make sense. The fact that agents must buy at least $\gamma_i$ of good $i$ will be responsible for making inequalities beneficial in certain regimes of productivity.

\subsection{A Primitive Economy}

Assume that every household works, either in firms of type 1 or in firms of type 2, such that $N_1 + N_2=N$, the total number of households. The budget $b$ of each household is spent either on good 1 or on good 2:
\begin{equation}
b = w = \sum_i c_i p_i
\end{equation}
Since $\theta_1 > \theta_2$, the optimal consumption schedule, as income grows, is to increase $c_1$ until $\Gamma_1$ while keeping $c_2=0$, beyond which $c_2$ starts picking up. More precisely, as long as $p_1 \gamma_1 \leq w \leq p_1 \Gamma_1$, one has:
\begin{equation}
c_1 = \frac{w}{p_1} = z_1; \qquad c_2 = 0.
\end{equation}
Setting $N c_1=Y_1$ leads to $N=N_1$, which is compatible with $N_2=0$. In this primitive economy, every household works to produce the basic subsistence good and good 2 is not produced at all. The total utility $\mathcal{U}$ is
\begin{equation}
\mathcal{U} = N \theta_1 z_1
\end{equation}
Now, as the productivity $z_1$ reaches $\Gamma_1$, households saturate their consumption of good 1 and the economy should be in a situation where good 2 starts to be produced. Since all households are assumed to have identical budgets and preferences, the total consumption of good 1 is, using market clearing and when $z_1 \geq \Gamma_1$:
\begin{equation}
C_1 = N \Gamma_1 = z_1 N_1 \quad  \longrightarrow \quad N_1 = N \xi_1,
\end{equation}
with 
\begin{equation}
\xi_1:=\frac{\Gamma_1}{z_1} \, <1
\end{equation}
the fraction of households working to produce good 1.

\subsection{A Threshold For Non-Essential Goods}

The household's budget available for good 2 is thus 
\begin{equation}
b_2 = w - \Gamma_1 p_1 = w (1 - \xi_1),
\end{equation}
such that consumption of good $2$ is
\begin{equation}
c_2 = \frac{b_2}{p_2} = (1 - \xi_1)z_2,
\end{equation}
at least as long as $c_2 \leq \Gamma_2$, i.e. when $1 \leq \xi_1 + \xi_2$. Market clearing for good 2 then yields:
\begin{equation}
Nc_2 = N_2 z_2 \quad  \longrightarrow \quad N_2= N(1 - \xi_1)
\end{equation}
which brings nothing new since we already know that $N_1=N \xi_1$. However, our assumption that goods come in indivisible quantum units, one should also have
\begin{equation}
c_2 \geq \gamma_2,
\end{equation}
or 
\begin{equation} \label{cond}
\frac{\gamma_2}{z_2} \leq 1 - \xi_1
\end{equation}
which is impossible if $z_2$ is smaller than a certain threshold $z_2^* := \gamma_2/(1-\xi_1)$.  Conversely, this can be stated as a threshold on $z_1$ for a fixed $z_2$: $z_1 < z_1^*$ with
\begin{equation}
z_1^* := z_2 \frac{\Gamma_1}{z_2 - {\gamma_2}}
\end{equation}
(and $z_1^*=+\infty$ when $z_2 \leq \gamma_2$).

In this case, good 2 cannot be produced and $N_2=0$. Within our over-simplified economy, the only solution that allows market clearing of good 1 and homogeneous wages is to assume that each household works part time, effectively reducing productivity as
\begin{equation}
z_1 \to z_1 \phi
\end{equation}
where $\phi \leq 1$ is the fraction of hours worked by households. The optimal total utility solution is thus
\begin{equation}
\phi = \xi_1.
\end{equation}
This is in agreement with intuition: if a single good can be produced, the total number of working hours decreases as production increases. The total utility is stuck to $\mathcal{U}= N \theta_1 \Gamma_1$, while extra utility from good 2 cannot be realized and remains latent. When $z_1 < z_1^*$, the problem is that nobody is rich enough to afford buying the ``luxury'' good, yet the basic good is produced so easily that full employment is not needed. 

The problem with such a solution is that any additional cost proportional to the number of employees (rather than to the number of working hours) would lead firms not to hire part time workers but rather to produce the same output with as few employees as possible. This would lead to massive unemployment, with a fraction $1- \xi_1$ of unemployed households unable to consume anything, and, in the absence of social subsidies, some surplus production -- clearly not a desirable situation.

In our model, increased productivity of essential goods thus leads to unemployment when wages are equal. This of course ties in with many arguments and narratives of the past. The Luddite movement is an obvious example, but even Keynes considered technological progress as a potential source of unemployment, and one of the underlying mechanism of the Great Depression (see the discussion in Robert Shiller's latest book, ``Narrative Economics'' \cite{shiller}). 

\section{A Two-Tier Economy}

A way out of this predicament is to allow some households to earn higher wages than others, kick starting the production of good 2. We will divide the households into $N_>$ high earners with wage $w_>$ and $N_<$ low earners with wage $w_<$, with $N = N_> + N_<$. We also assume for simplicity that the fraction of high earners is the same in the two types of firms, with the same average wage as before:
\begin{equation}\label{overlinew}
w= \frac{N_< w_< + N_> w_>}{N}.
\end{equation}
Therefore, we do not relate productivity to wages and skills, although this would lead to an interesting generalisation, with a different average wage in firms of type 1 and of type 2. In the present story, high earners might as well be chosen randomly with probability $N_>/N$. This is of course somewhat silly, but we want our model to be as bare-bone as possible, and refrain from adding any additional coupling between skills, productivity and wages.

Since the average wage is the same in all firms, prices are again obtained as:
\begin{equation}
p_i = \frac{w}{z_i}.
\end{equation}
We keep to the case where the basic good is easy to produce, i.e. $z_1 > \Gamma_1$ ($\xi_1 < 1$) henceforth. 

\subsection{Only High Earners Can Afford Non-Essential Goods}  

The first situation we shall consider is when low earners do not saturate consumption of good 1, whereas high earners do. In other words:
\begin{equation} 
\frac{w_<}{p_1} < \Gamma_1; \qquad \frac{w_>}{p_1} > \Gamma_1,
\end{equation}
or, using previous notations and $r_{<,>}:=w_{<,>}/w$ being the relative wages:
\begin{equation}
r_< <  \xi_1; \qquad r_> >  \xi_1.
\end{equation}
Total consumption of good 1 is thus
\begin{equation}
C_1 = {N_< r_< z_1} + N_> \Gamma_1 = N_1 z_1
\end{equation}
where the last equality holds because of market clearing. Hence
\begin{equation}\label{eq1}
N_1 = N_< r_< + N_> \xi_1.
\end{equation}
The high earners budget for the non-essential good 2 is
\begin{equation}
b_{2,>} = w_> - \Gamma_1 p_1 = w_> - \xi_1 w \quad  (> 0).
\end{equation}
Hence, consumption of good 2 is
\begin{equation}
C_2 = N_> \min[b_{2,>}/p_2,\Gamma_2].
\end{equation}
Clearly, high earners are overpaid if the consumption of good 2 is saturated, as this will lead to unemployment. So we assume
\begin{equation}
b_{2,>} \leq \Gamma_2 p_2 \quad  \longrightarrow \quad r_> \leq \xi_1 + \xi_2.
\end{equation}
In fact, the optimal value of $r_>$ will turn out to be equal to $\xi_1 + \gamma_2 \xi_2/\Gamma_2$, which satisfies this bound. 

Market clearing of good 2 then implies
\begin{equation}
N_> (r_> - \xi_1) z_2 = N_2 z_2 \quad  \longrightarrow \quad N_2 = N_> (r_> - \xi_1).
\end{equation}
Together with Eq. \eqref{eq1}, this yields:
\begin{equation}
N_1 + N_2 = N_< r_< + N_> \xi_1 + N_> (r_> - \xi_1)  = N
\end{equation}
where we have used Eq. \eqref{overlinew}. This means that the economic activity is maximized, reaching full employment. 

We still have to check that the condition $c_2 > \gamma_2$ can be fulfilled, so that production of good 2 can be viable. This requires:
\begin{equation}
b_{2,>} \geq \gamma_2 p_2 \quad  \longrightarrow \quad r_> \geq \xi_1 + \xi_2 \frac{\gamma_2}{\Gamma_2}.
\end{equation}

\subsection{Maximizing Utility and Minimizing Inequalities}

\begin{figure} 
  \centering
\includegraphics[width=.48\textwidth]{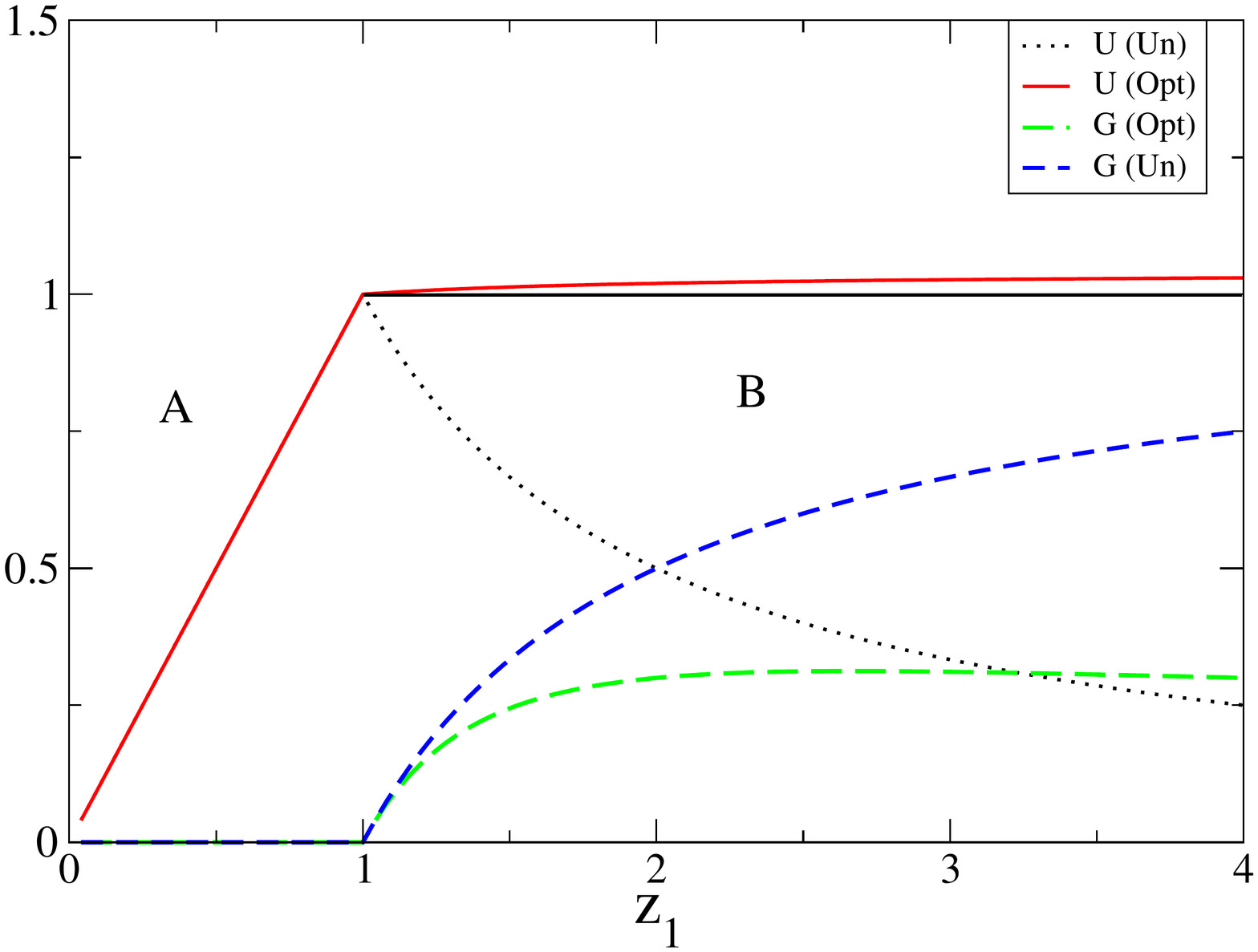}\, \includegraphics[width=.48\textwidth]{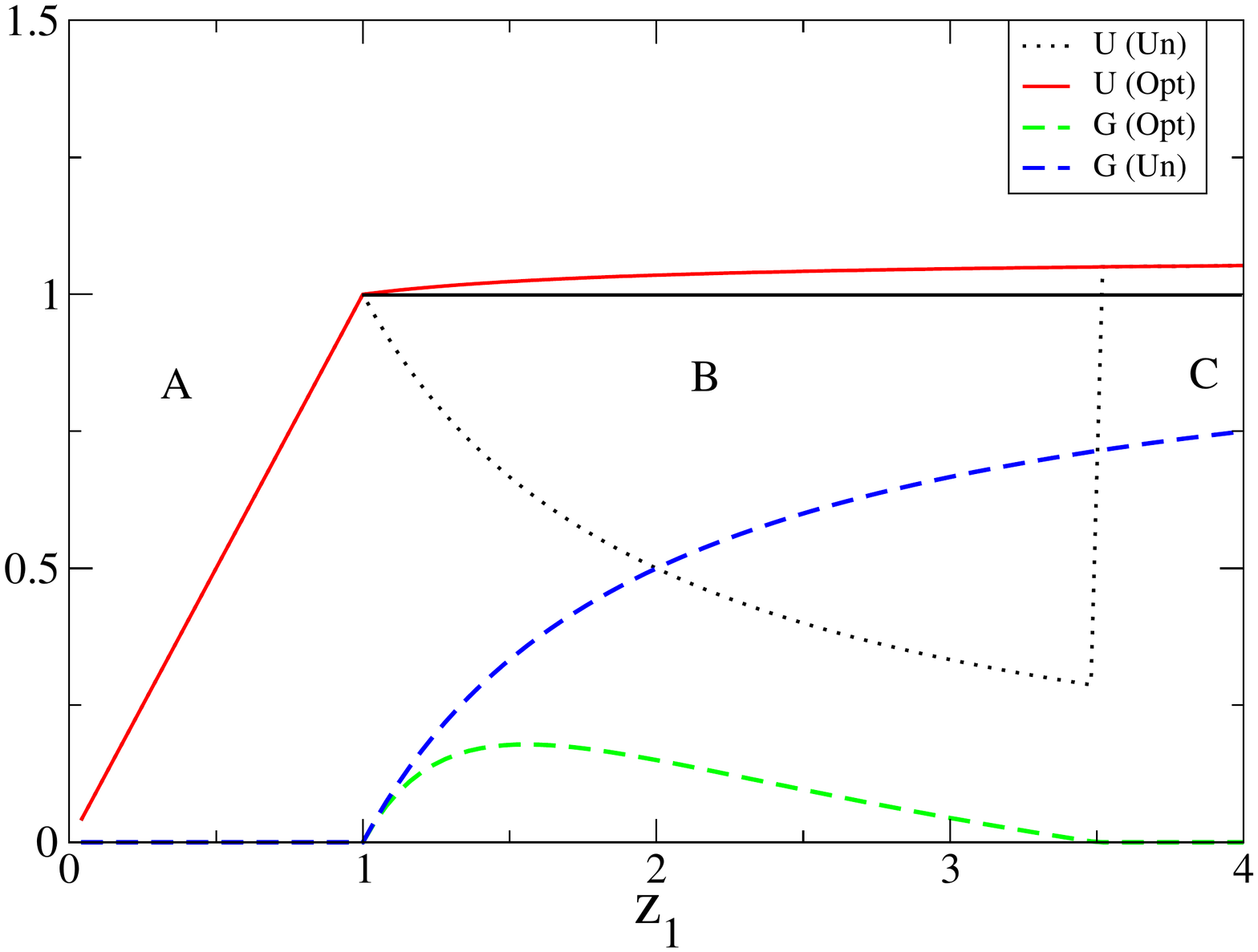}
  \caption{Utility per agent, $\mathcal{U}/N$, and Gini coefficients $G_{\min}$ (corresponding to optimal welfare) and $G_{\text{un}}$ (corresponding to partial unemployment), as a function of the basic good productivity $z_1$, for a fixed non-essential good productivity $z_2$ and for $\gamma_2=0.5$. Left: $z_2=0.4$, for which $z_1^*=+\infty$. Right: $z_2=0.7$, for which $z_1^*=3.5$. Other parameters are fixed to $\Gamma_1=1.$, $\theta_1=1.$, $\theta_2=0.1$. The black horizontal line corresponds to $\mathcal{U}/N=\theta_1 \Gamma_1$. The decaying dotted line corresponds to involuntary unemployment because good 2 is not produced, a situation that is curbed by unequal wages. When $z_2=0.7$, the production of good 2 becomes viable for $z_1 > z_{1c}=3.5$. Note that the Gini coefficient $G_{\min}$ is non-monotonic in that case. The letters A, B, C refer to the different phases in Fig. \ref{Fig2}.} 
  \label{Fig1}
\end{figure}

Introducing $f_{>,<}:=N_{>,<}/N$, we find that the total utility is given by
\begin{equation}
\frac{\mathcal{U}}{N} = (f_< r_< + f_> \xi_1) z_1 \theta_1 + f_> (r_> - \xi_1) z_2 \theta_2.
\end{equation}
Using $f_< + f_> = 1$ and $f_< r_< + f_> r_> =1$, this expression can be transformed into
\begin{equation}
\frac{\mathcal{U}}{N} = z_1 \theta_1 + \frac{(1- r_<)(\xi_1 - r_>)}{r_> - r_<}
(z_1 \theta_1 - z_2 \theta_2)
\end{equation}
Since $\mathcal{U}/N$ is an increasing function of $r_<$, a benevolent social planner would maximize this utility by setting $r_<$ to its maximum value
$\xi_1$, leading to 
\begin{equation}\label{max_ut}
\frac{\mathcal{U}}{N} = \Gamma_1 \theta_1 + (1- \xi_1) z_2 \theta_2,
\end{equation}
independently of $r_>$, which is not fixed by utility maximisation alone. The social planner might therefore look at the Gini coefficient, which is given by
\begin{equation}
G = \frac{(1-r_<)(r_>-1)}{r_> - r_<} = 1 - \xi_1 - \frac{(1 - \xi_1)^2}{r_> - \xi_1},  
\end{equation}
which is minimized for the smallest possible value of $r_>$. Hence the relative wages must be chosen as\footnote{Note that $r_>$ is indeed smaller than $\xi_1 + \xi_2$ since $\gamma_2 < \Gamma_2$.}  
\begin{equation}\label{min_eq}
r_< = \xi_1; \qquad r_> = \xi_1 + \frac{\gamma_2}{z_2},
\end{equation}
which corresponds to the minimum level of inequalities that achieves full employment and maximum welfare. The utility of low earners is not worse than in the homogeneous wage case with reduced working hours. The minimum value of the Gini coefficient is thus:
\begin{equation}\label{eq_gini}
G_{\min} = (1 - \xi_1)\left( 1 - \frac{z_2(1-\xi_1)}{\gamma_2}\right). 
\end{equation}
This is the central result of our paper. Note that, as expected, inequalities decrease as the productivity of good 2 increases. For a fixed $z_1$, $G_{\min}$ reaches zero when $z_2 \to z_2^*$: unsurprisingly, equal salary for all works again when technology of good 2 improves. For $z_2 < z_2^*$, partial unemployment would lead to a much worse total utility {\it and} Gini coefficient, given by $G_{\text{un}}=1 - \xi_1$, see Fig. \ref{Fig1}.

Note that the total utility, given by Eq. (\ref{max_ut}), continues to increase as $z_1$ increases past $\Gamma_1$, even when $z_1 < z_1^*$, contrarily to the homogeneous wage solution with reduced working hours, {\it a fortiori} to the solution with unemployment, see Fig. \ref{Fig1}. Note however that there is no disutility for labour in our model -- one could of course conclude that working less and only consuming basic goods is a better option, at the expense of leaving unexploited welfare for the rich on the table. This is, in a nutshell, the choice advocated by anti-capitalists. The value of $G_{\min}$ in Eq. \eqref{eq_gini} can otherwise be considered as a ``fair'' value for income inequalities, given the level of technology of the economy, in the sense that the global welfare is higher than if strict equality of wages was enforced. This in fact creates unemployment, i.e. another, perhaps worse, form of inequalities. On the other hand, Gini coefficients larger than $G_{\min}$ can be deemed unjustified and unfair.

\begin{figure} 
  \centering
\includegraphics[width=.7\textwidth]{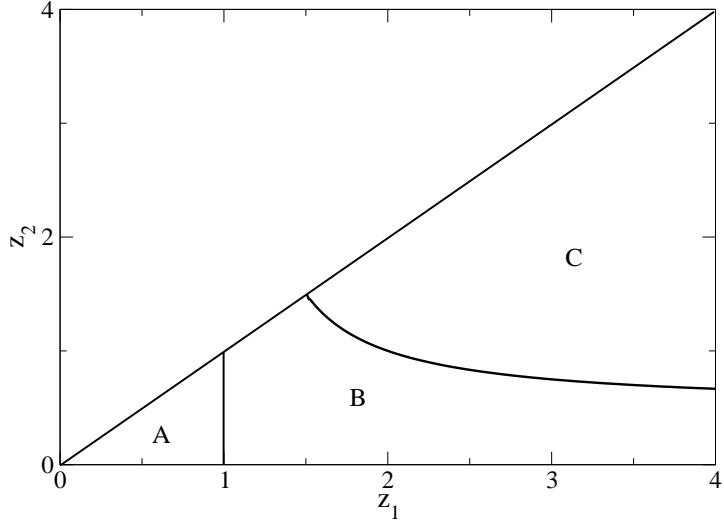}
  \caption{``Phase Diagram'' of the model, in the plane $z_1,z_2$, restricted to the region $z_2 < z_1$ where the non-essential good is more difficult to produce than the bare necessity good. Region A corresponds to a ``primitive economy'' where all wages are equal, but good 2 is not produced. Region B corresponds to a developing economy, where productivity of the essential good is moderately large ($\Gamma_1 < z_1 < z_1^*$, with $z_1^* := z_2 {\Gamma_1}/({z_2 - {\gamma_2}})$). Full employment can only be reached at the expense of wage inequalities. Region C, for large enough productivity $z_1 > z_1^*$, allows the economy to recover full employment with equal wages. Note that region B extends to $z_1 \to \infty$ when $z_2 < \gamma_2$.} 
  \label{Fig2}
\end{figure}

\subsection{Low Earners Saturate Consumption of Basic Goods}

Finally, let us study the case where productivity of good 1 is so high that both low earners and high earners can saturate their consumption of good 1, i.e. when
\begin{equation}
r_<, r_> >  \xi_1.
\end{equation}
The budget of low earners for good 2 is
\begin{equation}
b_{2,<} = w_< - \Gamma_1 p_1 = w_< - \xi_1 w \quad  (> 0).
\end{equation}
We keep the constraint that high earners do not saturate consumption of good 2, i.e.
\begin{equation}
b_{2,>} \leq \Gamma_2 p_2.
\end{equation}
Total consumption of good 1 is now
\begin{equation}
C_1 = N \Gamma_1 = z_1 N_1 \quad  \longrightarrow \quad N_1 = N \zeta_1.
\end{equation}
For low earners to buy good 2, their budget $b_{2,<}$ must exceed $\gamma_2 p_2$, or
\begin{equation}
r_< (1 - \xi_1) > \frac{\gamma_2}{z_2}.
\end{equation}
This is, again, only possible if $z_1 \geq z_1^*$, in which case total consumption of good 2 is
\begin{equation}
C_2 = \frac{1}{p_2} \, \left(N_< b_{2,<} + N_> b_{2,>}\right) = N_2 z_2
\end{equation}
or, using $p_2=w/z_2$
\begin{equation}
N_2 = N (1 - \zeta_1) \equiv N - N_1.
\end{equation}
So full employment is possible in this case, with $r_< = 1$, i.e. all wages equal -- we just recover the uniform case above. On the other hand, when $z_1 < z_1^*$ (or equivalently $z_2 < z_2^*$), low earners cannot buy good 2 and unemployment sets in. The different regimes of our model are summarized in Fig. \ref{Fig2}. 

\section{Conclusion}

We have tried to come up with the simplest possible economic model for which in some cases, wage inequalities lead to higher overall social welfare. In our story, this is due to the fact that the rich class can consume low productivity, expensive, non essential products, which allows everybody to remain employed even when the productivity of essential goods is high and producing them does not require everybody to work. In a sense, our model is a caricature of Tuscan XVth century Renaissance, when Florence's wealthy families started subsidising artists and painters.

Within our framework, growth and innovation appear to be intimately related to inequalities on two counts: (i) improved productivity of essential goods leads to unemployment when utility coming from consuming these goods saturates beyond a certain level, (ii) the production of low productivity, luxury goods can only take off if some high earners can afford them. 

In our bare-bone model, we find a relation between the minimum Gini coefficient required to make global welfare optimal and the level of technology, given by Eq. \eqref{eq_gini}, see also Fig. \ref{Fig2} and Eq. \eqref{gini2}. Stronger inequalities appear to be unjustified, both economically and in Rawles' sense. Conversely, if wage inequalities are not commensurate to the heterogeneity of productivity across products, total welfare is reduced. We note that while extra welfare is, in our model, captured by the high earners, the increase of economic activity should lead to increased tax revenues for the state, allowing part of this welfare to be redistributed. 

There are many directions in which our ``Tuscan Renaissance'' model should be extended to make it more realistic. One is to add some disutility of labor and a more elaborate labour market, where skills, productivity and wages are coupled. Another is to introduce not two, but a continuum of goods with different levels of productivity and utility, and relate the distribution of productivities to the optimal distribution of wages. Finally, our model suggests a specific beneficial impact of inequalities on growth by taking into account the fact that the appearance of a new good may be itself catalyze the emergence of more innovation, made possible by the production of that very good. 

We have focused above on income inequalities, but there might also be an optimal level of wealth inequalities that allows to kick start new activities in a more efficient way when only one investment decision is needed, rather than reaching a consensus within larger group of individuals, each of them unable to invest enough in the project to make it viable. 

The role of non-essential goods in the modern economy has in fact been put in a crude light by the Covid-19 pandemic. The debate over reverting to a leaner economy after the crisis -- where only ``essential'' goods are produced and wage inequalities are strongly reduced -- is already simmering.  We believe that such debates on income and wealth inequalities can only progress if we can come up with models that allow one to weigh both the beneficial and deleterious aspects of these inequalities, and normative prescriptions about an optimal level of inequalities with a clear notion of what exactly we want to optimise. The present paper is a modest attempt in that direction.   

\vskip 0.5cm

{\it Acknowledgments} I want to thank Michael Benzaquen, Leonard Bocquet, Beno\^\i t de Courson, Jos\'e Moran, David Thesmar and Francesco Zamponi for many discussions on these issues, and for most insightful remarks on the manuscript.

\appendix

\section*{Appendix: Non-Saturating, Concave Utility Functions} \label{sec:appendix}

Let us reformulate the main results of our paper in the case of a separable  constant relative risk aversion utility:
\begin{equation} \label{eqU2}
U = \sum_i \frac{\theta_i}{1-\sigma} c_i^{1-\sigma},
\end{equation}
with $0 < \sigma \leq 1$. For simplicity, we set $\sigma=1/2$ as a representative member of that family. Utility maximisation for a fixed budget $w$ leads to:
\begin{equation}
    c_i = z_i \psi_i; \qquad \psi_i:= \frac{z_i \theta_i^2}{\sum_i z_i \theta_i^2}.
\end{equation}
Using market clearing, one also has
\begin{equation}
    Y_i = N c_i = z_i N_i \to N_i = \psi_i N,
\end{equation}
and the whole workforce is fully employed. Such a solution is valid provided consumption of individual households is above the quantum of good $\gamma_i$. Again, this is not possible if the corresponding productivities are too low. In the case of two goods, when 
\begin{equation}
    c_2 = z_2 \psi_2  < \gamma_2
\end{equation}
the production of good 2 is stalled. But since the utility from good 1 always grows with consumption, households forego good 2 and instead spend all their budget on good 1:
\[
c_1 = z_1; \qquad c_2=0,
\]
still keeping full employment. The total utility is 
\begin{equation}
    \frac{\mathcal{U}_1}{2N}  := \theta_1 \sqrt{z_1}.
\end{equation}

Let us now repeat the calculations in this case, but with a fraction $f_{<,>}$ of the population earning $w_{<,>}:=r_{<,>} w$. When $z_2 \psi_2 r_< < \gamma_2$, low earners cannot afford good 2. Since the utility from good 1 always grows with consumption, these households spend all their budget on good 1:
\[
c_1^< = z_1 r_<;
\]
whereas for high earners
\[
c_1^> =  z_1 \psi_1 r_>; \qquad c_2^> =  z_2 \psi_2 r_>,
\]
with necessarily
\[
r_> \geq r_>^*:= \frac{\gamma_2}{z_2 \psi_2}.
\]
The labour market always clears since:
\[
Y_1 = N\left(f_< z_1 r_< + f_>  z_1 \psi_1 r_> \right) = z_1 N_1, 
\]
and
\[
Y_2 = N f_>  z_2 \psi_2 r_>  = z_2 N_2. 
\]
Hence:
\[
N_1 = N \left(f_< r_< + \psi_1 f_> r_>\right); \qquad N_2 = N \psi_2 f_> r_>.
\]
Using $\psi_1+\psi_2=1$ and $f_< r_< + f_> r_> = 1$, one finds $N_1+N_2=N$. 

The total utility is now:
\begin{equation}
    \frac{\mathcal{U}}{2N} = f_< \theta_1 \sqrt{z_1 r_<} + f_> \left(\theta_1 \sqrt{z_1 \psi_1 r_>} + \theta_2 \sqrt{z_2 \psi_2 r_>} \right).
\end{equation}
Since 
\[
f_< = \frac{r_> - 1}{r_> - r_<}; \qquad f_< = \frac{1 - r_<}{r_> - r_<}.
\]
the total utility is only a function of $r_<$ and $r_>$. 
So taking the derivative of $\mathcal{U}$ with respect to $r_<$ and $r_>$ leads to, respectively:
\[
\frac{\partial {U}}{\partial r_>} = \Xi_> \left[\sqrt{r_< r_>} - \frac{\beta}{2}(r_<+r_>)\right], 
\]
and
\[
\frac{\partial {U}}{\partial r_<} = \Xi_< \left[\frac{1}{2}(r_<+r_>)- \beta \sqrt{r_< r_>}\right], 
\]
where $\Xi_{<,>} \geq 0$ and 
\[
\beta:= \sqrt{\psi_1} + \frac{\theta_2 \sqrt{z_2 \psi_2}}{\theta_1 \sqrt{z_1}}.
\]
Since $r_<+r_> \geq 2 \sqrt{r_< r_>}$, one should separate two cases:
\begin{itemize}
    \item $\beta \leq 1$: in this case ${\partial {U}}/{\partial r_<}\geq 0$ and global welfare is maximized when $r_<=1$, and therefore $f_>=0$. A few high earners may exist, allow good 2 to be produced, but their number is so small that the utility is equal to  $\mathcal{U}_1$ and the Gini coefficient is zero.
    \item $\beta > 1$: the optimal low wage is now given by the equation
    \[
    r_<^2 + (2 - 4 \beta^2) r_> r_< + r_>^2 =0 \longrightarrow r_<=r_<^*:=2 \beta^2 - 1 - \sqrt{(2 \beta^2 - 1)^2 - r_>^2}.
    \]
    On the other hand, since ${\partial{U}}/{\partial r_>}\leq 0$, the optimal high wage is given by its lower bound:
    \[
    r_> = r_>^*:=\frac{\gamma_2}{z_2 \psi_2}
    \]
\end{itemize}
The last solution however only makes sense provided $r_<^* \leq 1$, i.e.
\[
\beta \geq  \frac{\sqrt{3 + r_>^{*2}}}{2} \geq 1.
\]
In this case, the Gini coefficient is 
\begin{equation}\label{gini2}
G = \frac{(1-r^*_<)(r^*_>-1)}{r^*_> - r^*_<}.
\end{equation}

Summarizing: for a non-saturating, concave utility function, one still finds that there is a region of parameters where inequalities do improve global welfare, when quantisation effects are taken into account. The phase diagram is however a little different from the one shown in Fig. \ref{Fig2}. For fixed values of $z_1$, $\theta_1$ and $\theta_2$, one finds that when $\gamma_2$ is below a certain threshold $\gamma_2^*$ (equal to $0.033/\theta_2^2$ when $z_1 \theta_1^2=1$), inequalities are never beneficial. When however $\gamma_2 > \gamma_2^*$, there exists a region $z_2 \in (z_2^\dagger, z_2^*)$ within which the optimal wage distribution has a positive Gini coefficient, i.e. when Eq. \eqref{gini2} leads to $G > 0$. 


\begin{thebibliography}{99}

\bibitem{Piketty1} Piketty, T. (2014). Capital in the 21st Century. (Cambridge, MA: Harvard University Press)

\bibitem{Piketty2} Piketty, T. (2020). Capital and Ideology. Cambridge: Harvard University Press.

\bibitem{Champ} Champernowne, D. G. (1953). A model of income distribution. The Economic Journal, 63(250), 318-351.

\bibitem{Angle} Angle, J. (1986). The surplus theory of social stratification and the size distribution of personal wealth. Social Forces, 65(2), 293-326.

\bibitem{chakraborti} Chakrabarti, B. K., Chakraborti, A., Chakravarty, S. R., \& Chatterjee, A. (2013). Econophysics of income and wealth distributions. Cambridge University Press.

\bibitem{boghossian} Boghosian, B. (2019). Is Inequality Inevitable? Wealth naturally trickles up in free-market economies, model suggests. Scientific American. 

\bibitem{brazil} Ribeiro, M. B. (2020). Income Distribution Dynamics of Economic Systems: An Econophysical Approach. Cambridge University Press.

\bibitem{Benhabib} Benhabib, J., Bisin, A., \& Zhu, S. (2011). The distribution of wealth and fiscal policy in economies with finitely lived agents. Econometrica, 79(1), 123-157.

\bibitem{Moll} Gabaix, X., Lasry, J. M., Lions, P. L., \& Moll, B. (2016). The dynamics of inequality. Econometrica, 84(6), 2071-2111.

\bibitem{imf} Ostry J. D., Berg A. \& Tsangarides C. G. 2014 Redistribution, Inequality and Growth (IMF Discussion note)
(Washington, DC: Intl Monetary Fund)

\bibitem{Rawles} Rawls, J. (2009). A Theory of Justice. Harvard university press.

\bibitem{Mankiw} Mankiw, N. G. (2013). Defending the one percent. Journal of Economic Perspectives, 27(3), 21-34.

\bibitem{Piketty} Piketty, T., \& Saez, E. (2012). A theory of optimal capital taxation (No. w17989). National Bureau of Economic Research.

\bibitem{bouchaud} Bouchaud, J. P. (2015). On growth-optimal tax rates and the issue of wealth inequalities. Journal of Statistical Mechanics: Theory and Experiment, 2015(11), P11011.

\bibitem{violante} Heathcote, J., Storesletten, K., \& Violante, G. L. (2017). Optimal tax progressivity: An analytical framework. The Quarterly Journal of Economics, 132(4), 1693-1754.

\bibitem{mazzucato} Mazzucato, M. (2013) The Entrepreneurial State. Debunking Public Versus Private Sector Myths (London: Anthem Press)

\bibitem{venka} Venkatasubramanian, V. (2017). How Much Inequality Is Fair?: Mathematical Principles of a Moral, Optimal, and Stable Capitalist Society. Columbia University Press.

\bibitem{shiller} Shiller, R. J. (2019). Narrative economics. Princeton University Press. 

\end{thebibliography}
\end{document}